\documentstyle[11pt,aaspp4]{article}
\received{}
\journalid{}{}
\articleid{}{}
\newcommand{\ts}{\thinspace}
\newdimen\sa  \def\sd{\sa=.1em  \ifmmode $\rlap{.}$''$\kern -\sa$
                               \else \rlap{.}$''$\kern -\sa\fi}
\newdimen\sb  \def\md{\sb=.03em \ifmmode $\kern \sb\rlap{.}$'$\null$
                                \else \kern \sb\rlap{.}$'$\fi}

\begin{document}
\title{{\it HST} ULTRAVIOLET SPECTRAL ENERGY 
       DISTRIBUTIONS \break FOR THREE ULTRALUMINOUS INFRARED GALAXIES\footnote{
       Based on observations with the NASA/ESA {\it Hubble Space Telescope},
       obtained at the Space Telescope Science Institute, which is operated 
       by AURA, Inc., under NASA Contract NAS 5-26555.}}

\author{Neil Trentham$^{\dagger}$, John Kormendy, D. B. Sanders}
\affil{Institute for Astronomy, University of Hawaii, 2680 Woodlawn Drive,
Honolulu, HI 96822,\\ E-mail: nat,{\ts}kormendy,{\ts}sanders@ifa.hawaii.edu}

\vskip 2pt
$^{\dagger}$current address: Institute of Astronomy, Madingley Road,
Cambridge CB3 0HA, United Kingdom

\message{STBASIC.TEX TeX Macro Library}
\message{ }

\def\cf{{\it cf.}}
\def\cp{{\it Call for Proposals\/}}
\def\ddd{$\ldots$}
\def\dddd{$\ldots$.}
\def\eg{{\it e.g.}}
\def\etal{{\it et~al.\ }}
\def\etc{etc.}
\def\ie{{\it i.e.}}
\def\ih{{\it Instrument Handbook}}
\def\ihs{{\it Instrument Handbooks}}
\def\instrucs{{\it Proposal Instructions\/}}
\def\iras{{\it IRAS\/}}        
\def\IRAS{{\it IRAS\/}}        
\def\iue{{\it IUE\/}}        
\def\IUE{{\it IUE\/}}        
\def\lk{Large\slash Key}
\def\mv{m_{_V}}
\def\Mv{M_{_V}}
\def\nl{{\it Newsletter\/}}
\def\nls{{\it Newsletters\/}}
\def\popi{Population~I}
\def\popii{Population~II}
\def\popiii{Population~III}
\def\pt{Panel\slash TAC}
\def\rprocess{{\it r-}process}
\def\sgch{subgiant CH star}
\def\sm{Small\slash Medium}
\def\sprocess{{\it s-}process}
\def\squig{\sim\!\!}
\def\subsun{_{\twelvesy\odot}}
\def\sun{\twelvesy\odot}
\def\Teff{T_{\rm eff}}
\def\vs{{\it vs.}}

\def\ArII{\hbox{Ar~$\scriptstyle\rm II$}}
\def\BaII{\hbox{Ba~$\scriptstyle\rm II$}}
\def\CI{\hbox{C~$\scriptstyle\rm I$}}
\def\CIII{\hbox{C~$\scriptstyle\rm III$}}
\def\CIV{\hbox{C~$\scriptstyle\rm IV$}}
\def\CaII{\hbox{Ca~$\scriptstyle\rm II$}}
\def\CrI{\hbox{Cr~$\scriptstyle\rm I$}}
\def\CrII{\hbox{Cr~$\scriptstyle\rm II$}}
\def\feh{$\rm [Fe/H]$}
\def\FeI{\hbox{Fe~$\scriptstyle\rm I$}}
\def\FeII{\hbox{Fe~$\scriptstyle\rm II$}}
\def\Halpha{H$\alpha$}
\def\Hbeta{H$\beta$}
\def\Hgamma{H$\gamma$}
\def\HI{\hbox{H~$\scriptstyle\rm I$}}
\def\HII{\hbox{H~$\scriptstyle\rm II$}}
\def\HeI{\hbox{He~$\scriptstyle\rm I$}}
\def\HeII{\hbox{He~$\scriptstyle\rm II$}}
\def\MgI{\hbox{Mg~$\scriptstyle\rm I$}}
\def\MgII{\hbox{Mg~$\scriptstyle\rm II$}}
\def\MnI{\hbox{Mn~$\scriptstyle\rm I$}}
\def\NI{\hbox{N~$\scriptstyle\rm I$}}
\def\NII{\hbox{N~$\scriptstyle\rm II$}}
\def\NIII{\hbox{N~$\scriptstyle\rm III$}}
\def\NV{\hbox{N~$\scriptstyle\rm V$}}
\def\NdII{\hbox{Nd~$\scriptstyle\rm II$}}
\def\NeI{\hbox{Ne~$\scriptstyle\rm I$}}
\def\NiI{\hbox{Ni~$\scriptstyle\rm I$}}
\def\OI{\hbox{O~$\scriptstyle\rm I$}}
\def\OIII{\hbox{O~$\scriptstyle\rm III$}}
\def\OV{\hbox{O~$\scriptstyle\rm V$}}
\def\OVI{\hbox{O~$\scriptstyle\rm VI$}}
\def\SI{\hbox{S~$\scriptstyle\rm I$}}
\def\SII{\hbox{S~$\scriptstyle\rm II$}}
\def\SiI{\hbox{Si~$\scriptstyle\rm I$}}
\def\SiII{\hbox{Si~$\scriptstyle\rm II$}}
\def\SiIV{\hbox{Si~$\scriptstyle\rm IV$}}
\def\SrII{\hbox{Sr~$\scriptstyle\rm II$}}
\def\TiI{\hbox{Ti~$\scriptstyle\rm I$}}
\def\TiII{\hbox{Ti~$\scriptstyle\rm II$}}
\def\YII{\hbox{Y~$\scriptstyle\rm II$}}

\def\ang{\AA}               
\def\deg{^\circ}
\def\degpoint{^\circ\mskip-7.0mu.\,}
\def\hm#1#2{$#1^{\rm h}#2^{\rm m}$}
\def\hms#1#2#3{$#1^{\rm h}#2^{\rm m}#3^{\rm s}$}
\def\kms{\>\rm km\>s^{-1}}
\def\minpoint{'\mskip-4.7mu.\mskip0.8mu}
\def\peryr{\>\rm yr^{-1}}
\def\secpoint{''\mskip-7.6mu.\,}
\def\spoint{^{\rm s}\mskip-7.0mu.\,}
\def\sqdeg{${\rm deg}^2$}

\def\spose#1{\hbox to 0pt{#1\hss}}
\def\lta{\mathrel{\spose{\lower 3pt\hbox{$\mathchar"218$}}
     \raise 2.0pt\hbox{$\mathchar"13C$}}}
\def\gta{\mathrel{\spose{\lower 3pt\hbox{$\mathchar"218$}}
     \raise 2.0pt\hbox{$\mathchar"13E$}}}
\def\la{Ly$\alpha$\ }
\def\ni{\noindent}
\def\in{\indent}
\def\inin{\in{\in}
\def\ininin{\inin{\in}}}
\def\hb{\hfil\break}

\begin{abstract}

      We present {\it HST\/} Faint Object Camera ultraviolet 
($\lambda$ = 2300 {\AA}, 1400 {\AA}) images of three
ultraluminous infrared galaxies (ULIGs: $L_{\rm ir} > 10^{12}$ 
$L_\odot$)\footnote{$L_{\rm ir} \equiv L(8-1000~\micron)$ computed from the 
flux in all four {\it IRAS\/} bands according to the prescription given in 
Perault (1987; see also Sanders \& Mirabel 1996).  Throughout this paper we 
assume that $H_0 = 75$ km s$^{-1}$ Mpc$^{-1}$ and that $\Omega_0 = 1$.} selected
from the $IRAS\/$ Revised Bright Galaxy Sample.  The purpose is to estimate 
spectral energy distributions (SEDs) to facilitate the identification of similar
objects at high redshift in deep optical, infrared, and
submillimeter surveys.  All three
galaxies, VII{\ts}Zw{\ts}031 (= IRAS F05081+7936), IRAS F12112+0305, and IRAS
F22491--1808,
were well detected with the F220W filter.  Two of the three were marginally 
detected with the F140W filter.  The fluxes, together with ground-based optical
and infrared photometry, are used to compute SEDs over a wide wavelength range.
The measured SEDs drop from the optical to the ultraviolet, but the magnitude 
of the drop ranges from a factor of $\sim 3$ in IRAS F22491$-$1808 to a factor
of $\sim 100$ in VII{\ts}Zw{\ts}031.  This is most likely due to different
internal 
extinctions.  Such an interpretation is also suggested by extrapolating to 
ultraviolet wavelengths the optical internal extinction measured in
VII{\ts}Zw{\ts}031.
$K$-corrections are calculated to determine the colors of the sample galaxies 
as seen at high redshifts.  
Galaxies like
VII{\ts}Zw{\ts}031 have very low observed rest-frame UV fluxes which means that 
such galaxies at high redshift will be extremely red or even missing in optical 
surveys.  On the other hand, 
galaxies like IRAS F12112+0305 and IRAS F22491--1808, if seen at high redshift, 
would be sufficiently blue that they would not easily be distinguished from 
normal field galaxies, and therefore, identified as ULIGs. 
The implication is then 
that submillimeter surveys may be the only means of properly identifying the 
majority of ULIGs at high redshift. 

\end{abstract}


\section{INTRODUCTION}

      Dissipation plays a major role during galaxy formation.  It may be the 
most important process that determines the fundamental properties of the visible
parts of galaxies (Blumenthal et al.~1984).  Ultraluminous infrared galaxies 
(ULIGs) are the best candidates we know about for galaxies that are currently
undergoing dissipative collapse (Kormendy \& Sanders 1992).  Their huge infrared
luminosities ($L_{\rm ir} > 10^{12}$ $L_{\odot}$) result from circumnuclear starbursts and/or 
active galactic nuclei (AGNs) surrounded by dust.  These galaxies contain large
amounts 
of molecular gas (e.{\ts}g., Sanders et al.~1991), with a large fraction of this
gas 
concentrated close to the nucleus -- typically $M_{\rm H_2} \sim 1 \times
10^{10}$ 
$M_\odot$ in the central 1{\ts}kpc diameter region (e.{\ts}g., Solomon et
al.~1997; 
Scoville et al.~1997) -- which presumably serves to fuel the nuclear activity.  
The morphologies of local ULIGs suggest
that the collapse was triggered by a recent merger (Joseph \& Wright 1985; 
Sanders et al.~1988a; Clements et al.~1996; Murphy et al.~1996).

      Kormendy \& Sanders (1992) argue that ULIGs may be giant elliptical 
galaxies in formation: they have solved the well-known problem (Ostriker 1980;
Carlberg 1986; Gunn 1987; Kormendy 1989, 1990) of producing the observed high 
core densities of ellipticals via an ongoing dissipative collapse and starburst.
Preliminary indications from the still-small number of ULIGs that have been
observed at high resolution suggest that they are consistent with the core 
fundamental plane correlations (e.{\ts}g., Kormendy \& Djorgovski 1989) of 
elliptical galaxies (Doyon et al.~1994).  It has also been argued that most
quasars and other powerful AGNs began their lives as ULIGs (Sanders et 
al.~1988a,b; for a recent review of these arguments, see Sanders \& Mirabel 
1996).  Therefore low-$z$ ULIGs, although rare, may provide an excellent 
opportunity to study the processes by which ellipticals and AGNs formed.  But 
before we can apply this hypothesis at the distances where most present-day
ellipticals formed and where quasars were most abundant, we need to be able to
find ULIGs at high redshifts.

      Ground-based followup observations of objects discovered by the {\it
Infrared Astronomical 
Satellite} ({\it IRAS\/}), and more recently by the {\it Infrared Space
Observatory}
have been productive in identifying ULIGs.  However, relatively few objects have
been identified from these surveys at $z > 1$.  For example, 
the hyperluminous galaxy IRAS F10214+4724 ($L_{\rm ir} \sim 10^{13}$ $L_\odot$; 
Rowan-Robinson et al.~1991) at $z = 2.32$, which is at least as luminous as the
most luminous
ULIGs observed locally and which is gravitationally lensed in the far infrared
by factors as high as 30 (Eisenhardt et al.~1996) only just made it into the 
{\it IRAS\/} 0.2 Jy sample (Oliver et al.~1996).  Until more sensitive 
far-infrared/submillimeter space telescopes are launched (e.{\ts}g. SIRTF and
FIRST), 
and prior to the successful operation of sensitive ground-based submillimeter 
detectors, the most productive method of searching for high-$z$ ULIGs (if they
exist) seemed to 
be from trying to identify heavily reddened sources in optical, and
near-infrared 
color and redshift surveys.  Such methods would rely heavily on comparing the 
colors of these candidate objects with the UV-optical colors of low-$z$ ULIGs. 
Within 
just the past year, an exciting new capability has been added with the
availability 
of new deep submillimeter observations obtained with the SCUBA camera on the
James Clerk Maxwell 
Telescope (JCMT) (e.{\ts}g.~Smail et al.~1997; Barger et al.~1998; Hughes et
al.~1998). 
The new deep submillimeter surveys are able to directly reveal which objects are
likely ULIG candidates, however optical and near-infrared data are still
required to 
obtain redshifts, and to determine more precisely the nature of the source of
the 
intense submillimeter emission.   An important issue will be to determine
if the objects found in these surveys by
virtue of their submillimeter luminosity are similar to local ULIGs.  If
this is the case, then we can use the detailed measurements of local ULIGs
as a template for studying the universe at high redshift.

      When we observe high redshift galaxies at optical and near-infrared
wavelengths, we are studying their rest-frame ultraviolet properties. 
Therefore,
in order to better identify high-$z$ ULIGs, we need to know the rest-frame
ultraviolet properties of local ULIGs. 
The ultraviolet colors of ULIGs are mostly determined by 
internal absorption.  Absorption can in principle produce arbitrarily red 
galaxies, although the reddest galaxies would also be faint.  It is not 
straightforward to model the color of a typical ULIGs as seen at high $z$.  
Ground-based optical surveys probe rest-frame ultraviolet wavelengths when $z 
\geq 1$.  To make color predictions, we would have to model the internal 
extinction of ULIGs at wavelengths where we have observations only from inside
our own Galaxy.  Its reddening properties are probably different from those of
typical ULIGs (see, e.{\ts}g.,~Kinney et al.~1994).  This clearly is dangerous.

      The purpose of this paper is to measure the ultraviolet colors (at $\sim$
2300 and $\sim$ 1400 \AA) of local ULIGs using the Faint Object Camera 
(FOC) and the {\it Hubble Space Telescope} ({\it HST\/}).  We will combine the
FOC 
measurements of these local ULIGs with ground-based optical and near-infrared
data to compute the 
SEDs for these galaxies.  We then compute $K$ corrections and hence predict the
colors of these galaxies as a function of redshift.  We can then compare these
colors with the observed colors of submillimeter-luminous galaxies found 
in the recent field surveys.

      Section 2 describes our sample and how it was selected.  Sections 3 and 4
present the {\it HST\/} and ground-based observations and data reduction, 
respectively.  In \S\ts5, we derive the SEDs and K corrections.  Section 6 
discusses the results and the applications to the
high-redshift Universe. 
Section 7 is a summary.

\section{THE SAMPLE}

      The basic properties of the galaxies we observed are listed in Table 1.
To get a meaningful average SED given the (probably different) amounts of 
extinction in different ULIGs, it is desirable to have as large a sample as
possible.  However,
ULIGs are faint in the ultraviolet, so we need to take deep exposures.  As a
result, we were able to observe only three objects.

      Our aim was to select an unbiased sample of ULIGs that might best be 
representative of their high-redshift counterparts.  IRAS F12112+0305 and
IRAS F22491$-$1808
were chosen from the $IRAS\/$ Revised Bright Galaxy Sample (Sanders et 
al.~1999) because they are typical ULIGs based on their ``cool'' 25 -- 60 
$\micron$ colors ($f_{25}/f_{60} < 0.3$; de Grijp et al.~1985; $f_{25}$ and 
$f_{60}$ are the {\it IRAS\/} flux densities in Jy at 25 and 60 $\micron$\,
respectively).  Most ULIGs at $z \simeq 0$ ($\sim${\ts}80{\ts\%) are cool.  
Rarer ``warm'' ULIGs like Mrk 231 which have strong evidence for a non-thermal
origin for the far infrared luminosity were avoided.  Our third source
VII{\ts}Zw{\ts}031 
has been studied in detail by Djorgovski et al.~(1990), who interpret it as a
merger-induced 
starburst galaxy.  Its {\it IRAS\/} luminosity is only 1\ts\% below the limiting
threshold of $10^{12}$ $L_\odot$ that Sanders et al.~(1988a) used to define
ULIGs; it therefore
was not in their ultraluminous sample (also it is at low Galactic latitude).  
However, we selected it here because it is one of the best examples of an {\it
IRAS\/} galaxy that appears to derive most of its infrared luminosity from an
embedded 
starburst.  We did not choose Arp 220, the nearest ``cool'' ULIG,
because it is too
big to fit in the small (11 arcseconds) field of view of the FOC.

      Ground-based R-band images of the ULIGs observed with the FOC are 
shown in Figure 1.  In VII{\ts}Zw{\ts}031, there
is no morphological evidence at large radii for an interaction, but at small 
radii, the isophotes deviate strongly from elliptical symmetry.  The other two
galaxies show much stronger evidence for a recent merger.  In both cases, the 
images suggest that the progenitors were disk galaxies.  The high infrared and
CO luminosities (Sage \& Solomon 1987, Sanders et al.~1991) suggest that the
progenitors were gas-rich.

\section{{\it HST\/} OBSERVATIONS AND DATA ANALYSIS}

      FOC images of all three galaxies were obtained using the F220W filter
($\lambda_{\rm peak} =$ 2270 \AA) and the F140W filter ($\lambda_{\rm peak} =$
1370 \AA).  Dates, exposure times, and measured fluxes (see below) are given in
Table 2.  Two or three exposures were taken of each galaxy with each filter. 
The offsets between them were found to be negligible (images of a single galaxy
were always taken sequentially), so the images were simply averaged.  All images
were recalibrated using the CALFOC package in STSDAS; the results were identical
to those produced by the {\it HST\/} pipeline calibration.  They show the usual
FOC blemishes: vertical stripes on the right side of the images due to 
variations in the camera scan speed and a fingerprint pattern in the 
geometrically corrected image.  Typical count rates in our images were 0.001 to
0.01 counts s$^{-1}$ pixel$^{-1}$; this is sufficiently low that we need not
worry about the nonlinearity of the FOC detector (see, e.{\ts}g., Baxter 1993).
In the rest of this section we describe the images and how we carried out the
photometry.

\subsection{\it VII{\ts}Zw{\ts}031 = IRAS F50081+7936}

      Our F220W image of VII{\ts}Zw{\ts}031 is shown in Figure 2.  The raw image
and
two deconvolutions using different techniques are shown.  The most striking 
feature is the bright crescent of knots in the deconvolved images.  The strong
similarity of the two deconvolutions argues that the morphology is not sensitive
to the deconvolution technique.  The knots are also visible in the raw image but
are less pronounced there.  They are not seen in optical ground-based images.
While the knots are interesting -- they may be luminous young star clusters --
we do not attempt to carry out photometry on the deconvolved images.  
This is because the deconvolutions do not conserve flux locally.  All
photometry is based on the raw images.

     Aperture brightnesses were measured by centering on the center of curvature
of the crescent of knots; this position is $< 1^{\prime\prime}$ from the optical
center of the galaxy.  The growth curve is shown in Figure 3.  The limiting
factor in determining the total brightness is the uncertainty in the sky level.
Determining the sky brightness directly is not possible because of the FOC 
blemishes.  Instead, we measure the sky value at various places well away from
the galaxy and plot growth curves for the range of resulting sky values.  Five
such curves are shown in Figure 3. We then identify which growth curve converges
best to a plateau as far as possible from the adopted center of the galaxy while
still indide the FOC field of view, 
and use it to estimate the total counts in the galaxy.  The other growth
curves provide an estimate of the uncertainties.  These are large: we
adopt a total flux of $9400 \pm 2000$ counts.  However, it will turn out that
the uncertainties due to sky subtraction are small compared to the differences
between different ULIGs.  Therefore sky subtraction uncertainties do not have a
major effect on our conclusions about the average properties of ULIGs.

      That the growth curves converge to a plateau at all suggests that the 
part of the galaxy that is significantly above the sky at 2300 {\AA} is well 
within our field of view.  However, it is possible that the structure we see
is embedded in a low-surface-brightness component that contributes significantly
to the total flux but that is never as bright as the rms sky noise.  Such a
halo would be excluded from the above total flux.  We estimate as follows how
important it could be.  We assume that this component has an exponential light
profile and a scale length of $h = 1$ kpc.  This is a typical scale length for 
a 0.5\ts$L^{*}$ galaxy (Freeman 1970; $L^*$ is the characteristic luminosity of
the Schechter 1976 luminosity function); 0.5\ts$L^{*}$ is probably close to the 
luminosity of the individual galaxies that merged to make VII{\ts}Zw{\ts}031
(Djorgovski
et al.~1990).  We then measure the rms sky noise $\sigma$ at various places in
the image and then run a detection algorithm (FOCAS, Jarvis \& Tyson 1981) to
look for objects with fluxes at least 3\ts$\sigma$ above the background and with
sizes at least as large as the seeing disk ($\sim$ 0\sd1).  When we do this and
find no objects, this sets a limit on the flux within one seeing disk of this 
exponential component.  Combining this limit with the scale length then allows
us to set a limit on the total flux in this component.  The additional 
luminosity that can be hidden in such a component is 1000 counts for the F220W 
image of VII{\ts}Zw{\ts}031.  This is much less than the measured counts.

      One complication is that the sky noise varies from place to place in the
image.  Also, the statistics are altered by the fingerprint pattern introduced
by the geometric transform.  However, $\sigma$ in regions with and without the
fingerprint differs by only $\sim$ 20\ts\%; the resulting uncertainty of $\pm$
200 counts is negligible compared to the uncertainty of $\pm$ 2000 counts due 
to the poorly known sky value.  Also, $\sigma$ varies from place to place by
$\sim$ 20\ts\% even in regions that do not show the fingerprint.  We therefore
just use a value of $\sigma$ that is averaged over a region large enough to have
substantial contributions from both fingerprint and non-fingerprint areas.  

      The calculations above suggest that we cannot hide a significant fraction
of the measured flux in a low-surface-brightness component that has a scale 
length of $h \simeq 1$ kpc.  However, if this component were more extended, the
amount of flux that could go undetected would increase rapidly with $h$ (it is 
approximately proportional to $h^2$).  A component with $h = 3$ kpc could 
dominate the total counts.  This is not an unreasonably large size;
the string of knots that we observe has a length of almost twice this.
What this means is that if we wish to compute total magnitudes, the values
we compute directly from the growth curve in Figure 3 might be substantially
in error.  Therefore we shall restrict ourselves to aperture magnitudes
here and in the rest of this paper.  We may still be missing some of the total
flux
in our apertures if there exists a low surface-brightness component, but
this is now a far smaller fraction of the total flux, even if this
component has $h = 3$ kpc. 

The above paragraph demonstrates that
we need to use aperture magnitudes.  We now 
need to select a physical size for the aperture in which to measure
fluxes.  This is chosen
to be 6 kpc in each galaxy rest frame,
which is the minimum size at which all the F220W growth curves
have converged to a plateau.  It is desirable to use the same
aperture for all the galaxies, since in the subsequent sections of the
paper we shall be comparing their SEDs.

      VII{\ts}Zw{\ts}031 was only detected at the 0.45$\sigma$ level in the
F140W images. 
These images are therefore best regarded as just giving flux limits.
A similar aperture analysis to that discussed above (see Figure 4)
gives a total flux of $250 \pm 550$ counts.  The contribution from a possible
low-surface-brightness component is proportionately much larger here than for 
the F220W data.  We estimate that the total counts that can be hidden in an 
$h = 1$ kpc component are 3700, more than 10 times the measured counts.

      For both images, the counts are then converted to units of flux density
$f_{\lambda}$ using the standard FOC calibration equations.  We assume that the
spectrum is flat across the bandpass (this assumption will be examined in detail
and corrected for in \S\ts5.4).  The resulting flux densities within the
6 kpc aperture are listed in the
final column of Table 2.  

\subsection{\it IRAS F12112+0305}

      The F220W image of IRAS F12112+0305 is displayed in Figure 5.  The
F220W-band emission is more diffuse than it is in VII{\ts}Zw{\ts}031; there is
no 
evidence here for luminous star clusters.  The emission comes from two 
concentrations; neither is centered near the optical centers of the interacting
galaxies.  Their center positions are shown in Figure 5; they are close to the
edge of the field of view of the detector.  Clearly we are more likely to miss
flux at large radii than we were for VII{\ts}Zw{\ts}031.  In Section 5.3, we
discuss how
such missing flux would affect the calculation of the SED and the $K$
corrections.

      Figure 6 shows growth curves given by aperture photometry like that in
\S\ts3.1.  We centered the apertures mid way between the two light 
concentrations.  Adopting the sky value that results in a plateau in the growth
curves implies a flux of $15,000 \pm 2,000$ counts.  

      The F140W images show faint extended emission distributed like the
emission in the F220W images.  A growth curve analysis like the above (see
Fig.~7) gives a measured flux of $6000 \pm 2000$ counts.  

      The counts were then converted to physical flux units; the results are
presented in Table 2.  As for the previous galaxy, these refer to 
6 kpc aperture magnitudes.   

\subsection{\it IRAS F22491-1808}

      The F220W and F140W
images of IRAS F22491$-$1808 are shown in Figure 8.  We detect
this galaxy at a far higher level of significance than the other two galaxies.
The integration time was lower by a factor of $\sim 10$, yet we have
approximately the same number of counts.  There is some evidence for knots, but
they contribute a substantially smaller fraction of the 2300 {\AA} flux than do
the knots in VII{\ts}Zw{\ts}031.  Figure 8 also shows that the brightest peak in
the 
F220W image is offset from the optical center by about 2$^{\prime\prime}$.  We
regard it as unlikely that such a large ($> 5 \sigma$) peak at 2300 {\AA} would
be undetected in our $B$-band optical image.  We therefore attribute the offset
to errors in {\it HST\/} pointing (errors this large are not uncommon -- see 
the FOC Data Handbook, Baum 1994).  There is further evidence for this
interpretation: if we overlay the F220W peak on the optical peak, then the faint
emission that we see in the F220W image to the east of the main galaxy exactly 
coincides with the eastern extension in the optical image (see Figure 1).

      The photometry in this case is complicated because the galaxy falls close
enough to the west edge of the image that variations in the camera scan speed 
cause substantial variations in the background in the F140W image
(see Figure 8b).  
However, these variations have 
a definite direction (up-down in the image).  We therefore perform photometry 
not using circular apertures but using rectangular apertures, where we vary the
aperture dimension in the direction orthogonal to the fluctuations.  When we
subtract the sky, these fluctuations will affect the subtraction in a systematic
way, so that the mean sky values we subtract which then produce a plateau in the
growth curve will be slightly different from the mean sky values we measure in 
the main part of the image.  These effects are not large ($< 10$\ts\%), and
they are easily corrected, as above.  We then find that the growth curves 
(Figure 9) converge to a plateau; this also suggests that the measured flux
above sky drops to zero around the edges of the image.  The total flux is $9300
\pm 1000$ counts.  
The F140W image shows a single brightness peak at the same place as the
brightest knot in the F220W image and a very faint diffuse component that also
follows the emission in the F220W image.  Aperture photometry gives a total
flux of $1500 \pm 700$ counts.  

Table 2 again lists the fluxes in physical units.

\section{GROUND-BASED OBSERVATIONS AND DATA ANALYSIS}

      To compute SEDs and $K$ corrections as input for the identification of
ULIGs in deep surveys, we need not only ultraviolet measurements but also 
optical and near-infrared photometry.  We obtained this mostly from ground-based
observations summarized in Table 3.  We also used some images taken by other
observers and photometry from the literature.  All of our observations were 
taken at the University of Hawaii 2.2 m telescope.  The detector was a thinned
Tektronix 2048 $\times$ 2048 CCD for the optical images and the QUIRC 1024
$\times$ 1024 HgCdTe array for the near-infrared images.  The field of view is
7\md5$\times$7\md5 for the CCD observations at the f/10 focus, 3\md2$\times$3md2
for the 
QUIRC observations at f/10, and 1\md0$\times$1md0 for the QUIRC observations at
f/31.
These fields were easily large enough to contain all of the galaxy light.  We 
used standard data reduction techniques for our images; twilight sky flats were
used for the optical images and dome flats for the near-infrared images.  Most 
of the optical images were taken as three short exposures; at least six were 
taken for the QUIRC images.  Cosmic rays and bad pixels in the vicinity of the
galaxy were then identified by their appearance in only one of the individual 
images; they were removed by hand. Instrumental magnitudes were computed from 
observations of 20 -- 30 standard stars, and the photometry was converted to 
the Johnson $UBV$ -- Cousins $RI$ magnitude system of Landolt (1992).  For the 
$K^\prime$ observations, the typical number of standards observed per night was
6, and we used the color conversions of Wainscoat \& Cowie (1992) to convert to
the magnitude system of Elias et al.~(1982).  The zero points are accurate to
1.5\ts\% in the optical and 3\ts\% in the near-infrared.

      We then measured the 
magnitudes of the galaxies within an aperture that corresponds to 6 kpc
in the rest frame of each galaxy, as outlined in the previous section.
This alignment was performed as follows.
For VII{\ts}Zw{\ts}031 we centered the optical galaxy at the center of curvature
of the
knots -- this is also the position we would choose if we were to assume
that the $HST$ pointing was completely accurate.  For IRAS F22491$-$1808,
we aligned the images by requiring the UV and optical peaks to coincide,
as outlined in Section 3.3.  
For IRAS F12112+0305, the aperture was centered 
mid way between the UV peaks (see Section 3.2), 
which is approximately half way between the optical
peaks (see Figure 5).  Adopting these centers, we found that the
fraction of the light above the 3$\sigma$ isophote
falling outside the aperture is about 20\% or less.  This is small
enough that highly accurate alignment between the optical and
UV images is not essential. 
For these bright galaxies, the main source of error 
in the magnitudes is the then uncertainty in the zero point.   The aperture
magnitudes are presented in Table 3.  

      For IRAS F12112+0305 and IRAS F22491--1808, we obtained several exposures
in a number of passbands (Table 3).  Our $B$-band image of IRAS F12112+0305
taken in 1995 May revealed a bright, point-like object in the southern tidal
feature 
that did not appear in any of the other images.  This is probably a 
supernova (Trentham 1997).  This fell close to the edge of our
aperture (it was 7\sd7 = 10 kpc from the galaxy center). 
We measured the total flux in the supernova using
DAOPHOT (Stetson 1987) and subtracted the part of it that fell within
the detection aperture to derive the magnitude presented in Table 3.
The supernova had a total $B$
magnitude of 18.83 corresponding to an absolute magnitude of $M_B = -18.55$ if
it is in the galaxy.  This supports the interpretation as a supernova and
suggests that it was observed close to its maximum luminosity (Trentham 1997). 

      For VII{\ts}Zw{\ts}031, we had no photometric images of our own, so we
measured the
total magnitude on a $K$-band image from 
Mazzarella et al.~(1999).  Optical 
measurements in the $gri$ filter system (Thuan \& Gunn 1976; Wade et al.~1979)
were taken from Djorgovski et al.~(1990).

\section{SPECTRAL ENERGY DISTRIBUTIONS AND K CORRECTIONS}

      In this section we use the results of the previous two sections to compute
SEDs for the three galaxies.  We then compute $K$ corrections; these are defined
as follows. 

      Suppose we observe the light from a galaxy of redshift $z$ and absolute
magnitude $M_{\rm X}$ in some filter X.  Its apparent magnitude $m_{\rm X} (z)$
is
$$m_{\rm X} (z) = M_{\rm X} + \mu(z) + K_{\rm X} (z).               \eqno(1)$$
Here $\mu(z)$ is the distance modulus, and $K_{\rm X}(z)$ is the $K$ correction.
It corrects for two effects: (i) the redshifted spectrum is stretched through
the bandwidth of the filter, and (ii) the rest-frame galaxy light that we see 
through the filter comes from a bluer part of the SED because of the redshift.
The second effect is normally the larger, depending on the SED.  If $K(z)$ is
known for filters X and Y, then we can predict the colors as a function of
redshift:
$$m_{\rm X} (z) - m_{\rm Y} (z) = M_{\rm X} - M_{\rm Y} 
                                + K_{\rm X} (z) - K_{\rm Y} (z).    \eqno(2)$$
Candidate ULIGs can then be identified in photometric surveys on the basis of 
color information.

The $K$ correction for filter X is
$$K_{\rm X} (z) = 2.5 \log_{10} \left[(1+z) 
                  { {\int_0^{\infty} f_{\lambda} (\lambda^{\prime}) 
                  T_{\rm X} (\lambda^{\prime}) {\rm d}{\lambda^{\prime}}
                  }\over{\int_0^{\infty} f_{\lambda} 
                  ({{\lambda^{\prime}}\over{1+z}}) T_{\rm X} (\lambda^{\prime}) 
                  {\rm d}{\lambda^{\prime}}}}\right]                \eqno(3)$$ 
(e.{\ts}g., Sandage 1995).  Here $f_{\lambda}$ is the flux density of the 
source, and $T_{\rm X} (\lambda)$ is the sensitivity of the detector plus filter
X.  In this work, we measure $f_{\lambda}$.  We can then calculate $K$
corrections for these galaxies as functions of $z$ for various filters X.  A 
common simplification that is used when the data are poorly sampled (e.{\ts}g.,
if only broadband colors are available, as in this paper) is to assume that the
filter X is narrow; then $T_{\rm X} (\lambda)$ is a heavily peaked function of
wavelength $\lambda$ centered on some reference wavelength $\lambda_{0}$.  That
is, {$T_{\rm X} (\lambda) \approx \delta (\lambda - \lambda_{0}$).  Then 
equation (3) simplifies to
$$K_{\rm X}(z) \approx K_{\lambda_{0}}(z) =  2.5 \log_{10}\left[{ 
            (1+z) { {f_{\lambda} (\lambda_{0})}\over{f_{\lambda}(
            {{\lambda_{0}}\over{1+z}})} }}\right] .                 \eqno(4)$$
Equation (4) is also valid if the SED is flat across the bandpass.  We shall use
it; the optical filters for which we compute $K$ corrections are narrow enough 
that the uncertainties introduced by using equation (4) instead of equation (3) 
are much smaller than the errors inherent in computing $f_{\lambda}$ from the 
{\it HST\/} observations, where the filters are not narrow.  In \S\ts5.4, we 
show how the calculation of $f_{\lambda}$ from broad-band colors (as in \S\ts3) 
requires us to assume a functional form for $f_{\lambda}$ within the bandpasses,
and we show how this functional form affects the SED.  (In \S\ts3, we assumed a
flat spectrum, $\alpha = 0$, where $f_{\lambda} \propto \lambda^{\alpha}$.)  We
then show how the functional form affects the computation of $K$ corrections
using equation (4).  We find that assuming a functional form results in
compensating systematic errors when the SEDs are calculated and then when the
$K$ corrections are calculated from the SEDs. 
We also show in \S\ts5.4 
that the magnitude of the errors made in using equation (4) are negligible 
compared to the differences between ULIGs.  They are therefore not important 
when we wish to decide whether the objects that we see in color surveys are 
ULIGs or not.

      In the rest of this section, we examine the assumptions and corrections
we need to make to compute SEDs and $K$ corrections.  The results are presented
and discussed in the next section.

\subsection{\it Distance Scale}

      We assume a standard Friedmann cosmology with a Hubble constant of $H_0 =
75$ km s$^{-1}$ Mpc$^{-1}$ and a cosmological density parameter of $\Omega_0 =
1$.  Then fluxes are related to luminosities as
$$L_\lambda = 4 \pi d_{L}^{2}(z) f_\lambda.                         \eqno(5)$$
Here $L_\lambda$ is the luminosity density (the bolometric luminosity is $L =
\int_0^{\infty} L_{\lambda} {\rm d}{\lambda}$), and $d_{L} (z)$ is the 
luminosity distance.  For $\Omega_0 = 1$,
$$d_{L} (z) = { {2 c}\over{H_0}} ( 1 + z - (1 + z)^{{1\over2}}).    \eqno(6)$$ 

\subsection{\it Galactic Extinction}

      The fluxes and magnitudes quoted in \S\S\ts3 and 4 are the observed 
values; they need to be corrected for Galactic extinction.  We adopt $B$-band 
Galactic extinctions from Burstein \& Heiles (1982); they are listed in Table 1.
To find the extinctions in other optical bandpasses, we use the color 
conversions of Cardelli et al.~(1989).  At ultraviolet wavelengths, we used the 
extinction curves given by Mathis (1990). For VII{\ts}Zw{\ts}031, the Galactic
extinction
is quite large (Djorgovski et al.~1990).  For $A_B = 0.36$, the extinction in 
the F220W filter is $A_{2270~\AA} \approx 0.84$.  The extinction at this
wavelength is particularly large because of the 2200 {\AA} graphite bump 
(Mathis 1990).  We apply the Galactic absorption corrections to the observed
fluxes or magnitudes before computing the luminosity using equation (5).  Note
that these corrections are {\it much smaller} than the corrections we would have
to make for internal reddening (see Djorgovski et al.~1990) if we wanted to 
determine the intrinsic luminosities of the galaxies at optical wavelengths.  
Since our aim is to compare local and high-$z$ ULIGs, we do not attempt to 
correct for internal extinction. 

\subsection{\it Spatial Distribution of Flux} 

The FOC field of view is only 11\sd3$\times$11\sd3.  This corresponds to 
10.9 kpc for VII{\ts}Zw{\ts}031, 14.0 kpc for IRAS 
F12112+0305, and 14.9 kpc for IRAS F22491$-$1808.  Most of the light of normal
galaxies is contained within these diameters, but it is not clear that the same
is
true for ULIGs.  Large amounts of internal extinction may reduce the fraction of
the ultraviolet light that is near the galaxy center.  
This was in essence our motivation for using aperture magnitudes.

      For all of our images, the growth curves converged on constant total
brightnesses well within the FOC field of view.  While this is encouraging, the
caveat is that we determined sky values within the same images, so we might 
have subtracted a low-surface-brightness component that extends beyond the
edges of the FOC images.  Ultraviolet observations with a larger field of view
would be required to measure this light or show that it is unimportant.  
meanwhile, we caution the reader that the 
ultraviolet parts of the SEDs may be too faint and therefore that the true $K$
corrections may be smaller than the ones we compute at wavelengths $\lambda$ 
and redshifts $z$ such that $(1+z)^{-1} \lambda < 4400$ {\AA}.  Longward of 
4400 {\AA}, the $K$ corrections are determined only by the optical measurements.

       One other complication is that the IRAS beam 
($2^\prime \times 4^\prime$) is much larger than
the field of view of the FOC -- therefore when we place far-infrared and
UV detections on the same SED, we need to be aware of the fact that
we might be probing spatially distinct regions of the galaxy.   
This problem only disappears if 
the far-infrared emission from ULIGs comes from a very compact
region (which is highly likely; see e.g.~Lonsdale et al.~1994 for Arp 220) that
is the same
region from which the UV flux comes. 
In the rest of this work, we only consider parts of the SED blueward of
the rest-frame $K$-band, so we can safely ignore this effect. 

 \subsection{\it The Slope of the SED Within the Bandpass}

  Each magnitude measurement is an integral of the flux density over some filter
bandpass, $\int_0^{\infty} f_{\lambda} T_{\rm X}(\lambda) {\rm d}\lambda$.  To
get the SED, we need the flux density $f_{\lambda}$ at some reference wavelength
$\lambda_0$; we choose this to be the peak transmission wavelength of the 
filter-telescope-detector system.  Therefore, we need to adopt some functional
form of $f_{\lambda}$ within the bandpass.  We assume a power law $f_{\lambda}
\propto \lambda^{\alpha}$. The problem is that $\alpha$ is unknown.  Crude 
estimates can be obtained by using the adjacent points to measure the 
approximate slopes of the SED on either side of the point in question.  However,
these estimates are not necessarily useful, (i) because the blueward and redward
point often give very different slopes, and (ii) because the distance between 
points in wavelength space is much bigger that the bandwidth of the filters.  A
ore quantitative analysis is suggested, as follows.  Suppose that we assume 
some value of $\alpha = \alpha_1$ when we compute our SEDs using the total
measured flux in filter X with reference wavelength $\lambda_{0,{\rm X}}$ and
that we do this incorrectly because the true value is $\alpha_{2}$.  Our SED 
point will then be incorrect by a factor
${ {I_{\rm X} (\alpha_1)}\over{I_{\rm X} (\alpha_2)}}$, where 
$$I_{\rm X} (\alpha) = { {\int_{0}^{\infty} \left({ {\lambda}\over
                         {\lambda_{0,{\rm X}}}}\right)^{\alpha} 
                         T_{\rm X} (\lambda) {\rm d}\lambda}
                         \over{\int_{0}^{\infty} T_{\rm X} (\lambda) 
                         {\rm d}\lambda}}.                         \eqno (7)$$
Now suppose that we compute the $K$ correction for a filter Y using the 
incorrect SED that we just derived.  Filter Y has a redder $\lambda_0$ than 
filter X because of the redshift $z$: $(1 + z) = 
{{\lambda_{0,{\rm Y}}}\over{\lambda_{0,{\rm X}}}}$.  The denominator of the term
inside the brackets in equation (3) will now be incorrect by a factor of
$f = { {I_{\rm X} (\alpha_1)}\over{I_{\rm X} (\alpha_2)}}
     { {I_{\rm Y} (\alpha_2)}\over{I_{\rm Y} (\alpha_1)}}$, and the $K$
correction will be incorrect by a factor of $2.5 \log_{10} f$.  (We assume that
the numerator of the term inside the brackets in equation (3) has been computed
correctly.)  Therefore, if the transmission curves for X and Y have the same 
shape, then $f = 1$ and the $K$ corrections will be independant of $\alpha$. 
This is almost true for the optical filters.  We therefore set $\alpha = 0$ 
when we compute the SEDs for the optical and K-band filters. For the {\it HST\/}
filters, complications are introduced by the broad wings of the F220W and F140W
transmission curves from $\lambda_{0}$ toward red wavelengths.  Our approach is
to assume $\alpha = 0$ here, too, when we compute the SEDs.  This value of
$\alpha$ is suggested by the IRAS F12112+0305 and IRAS F22491$-$1808 SED slopes
between the F220W and F140W point.  Doing this also has the advantage that the
SED is determined consistently for our optical and ultraviolet points.

      However, we still need a quantitative measure of how the assumed $\alpha$
affects our results.  We therefore present is Table 4 the function $I (\alpha)$ 
for the {\it HST\/} filters and for a typical optical filter.  We can then 
calculate the $f$ for any particular $\alpha$ and thereby determine how much the
$K$ correction is affected.  For the F140W filter, these numbers can be large.
The reason is the broad peak in the F140W transmission curve redward of the 
reference wavelength.  For example, at 6000 {\AA}, the F140W filter lets in 
twice as much light as the F220W filter.  However, the SEDs suggest that 
$\alpha \simeq 0$ for this point.  For the F220W filter, $\alpha$ is less likely
to be 0, because the SED in all cases drops strongly between the bluest optical
measurement and the F220W data point (although some of this is due to the 4000 
{\AA} break).  However, $I (\alpha)$ is much closer to 1 for the F220W filter
because it's transmission curve is narrower.  We investigated a number of
possibilites; typical errors introduced by our assumption of $\alpha = 0$ given
plausible values of $\alpha$ range from zero to a few percent.  The largest is
for $K$ corrections which use the F220W point of VII{\ts}Zw{\ts}031.  Here, the
true 
$\alpha$ may be as large as 4, based on the relative positions of the F220W 
point and  the $g$-band point (this is the bluest optical point in the SED).
This would mean that $f$ is about 16\ts\%.  While this is significant, it is 
smaller than the uncertainty in the measured counts and the differences in SEDs
between the galaxies.  Therefore the $K$ corrections are accurate enough to 
permit identification of ULIGs in photometric surveys.

      A potentially more serious worry is the contamination of the broadband 
colors by emission lines.  However, Figure 6 of Djorgovski et al.~(1990) 
suggests that the optical broadband colors of VII{\ts}Zw{\ts}031 are not
dominated by 
contributions from the lines.  Our FOC filters were chosen to be free of known
emission lines.  We therefore do not expect emission lines to be a problem.

      The SEDs are presented in Figures 11 -- 13 and in Table 5.  The plots show
$\nu L_{\nu} = \lambda L_{\lambda}$ as a function of frequency $\nu$; they are
computed assuming $\alpha = 0$.  The 2270 {\AA} and 1370 {\AA} points derived
from the {\it HST\/} photometry are shown as filled circles.  For
VII{\ts}Zw{\ts}031, 
the 1370 {\AA} point is a limit (i.{\ts}e., a 0.45 $\sigma$ detection).  

      From the SEDs, we calculate $K$ corrections $K_{\rm X} (z)$ for various
filters X with reference wavelength $\lambda_{0}$, where $\lambda_{0} 
(1+z)^{-1} > 1370$ {\AA} for IRAS F12112+0305 and IRAS F22491$-$1808, and where
$\lambda_{0} (1+z)^{-1} > 2270$ {\AA} for VII{\ts}Zw{\ts}031.  These $K$
corrections are
presented in Table 6.

\section{DISCUSSION}

      In this section, we first compare the SEDs of the three sample galaxies
and we suggest a physical interpretation.  We then 
discuss the implication of our results for 
ground-based submillimeter, near-infrared, and
optical color and redshift surveys.

\subsection{\it The SEDs of the Sample Galaxies} 

      The SEDs of all three galaxies drop between optical and ultraviolet 
wavelengths, but by substantially different amounts (see Figure 14).  
VII{\ts}Zw{\ts}031 has the steepest
drop; the SED drops be two orders of magnitude between the $g$ and F220W bands. 
IRAS F22491$-$1808 has the shallowest drop, i.{\ts}e., by less than a factor of
3 between the $B$ and F220W bands.  The straightforward interpretation is that
the three galaxies have different amounts of internal extinction from dust.  At
optical wavelengths, the internal extinction in VII{\ts}Zw{\ts}031 is $E_{B-V}
\simeq
0.9$ (Djorgovski et al.~1990).  This suggests that the internal extinction at 
far-ultraviolet ($\lambda < 3000$\AA)  
wavelengths should be huge (factors of hundreds are plausible given
typical extinction laws).  The extremely low ultraviolet flux that we observe is
therefore not surprising.  The ultraviolet light from the other galaxies is 
presumably less absorbed.  It is interesting that these three galaxies should
show such different properties, given that they were selected in an unbiased way
so as to be typical of their class.  

      Note also that VII{\ts}Zw{\ts}031, which has the most steeply dropping
ultraviolet
SED, has a substantial population of candidate young star clusters in the F220W 
image.  Their ultraviolet luminosities are $\approx 10^8$ L$_{\odot}$,
comparable 
to or brighter than the luminosities of the super star clusters seen in other
merging and starburst galaxies (e.{\ts}g., Whitmore et al.~1993; Conti \& Vacca
1994; Shaya et al.~1994; Whitmore \& Schweizer 1995; O'Connell et al.~1995;
Meurer et al.~1995; Schweizer 1996; Schweizer et al.~1996; Tacconi-Garman, 
Sternberg, \& Eckart 1997). The clusters are several kpc from the galaxy center;
we presumably see them 
because the optical depth is low there.  But they contribute much of
the ultraviolet light.  Therefore ULIGs with the internal extinction properties
of VII{\ts}Zw{\ts}031 but with no such clusters will have SEDs that drop even
more 
steeply in the ultraviolet than does the SED of VII{\ts}Zw{\ts}031.  Seen at
high 
redshifts, they would be even redder than VII{\ts}Zw{\ts}031.

\subsection{\it Application to High-Redshift} 

     In the past three years there have been a number of important breakthroughs
in the detection and identification of galaxies with redshifts $ 1 < z < 4$.    
Significant new populations include 
Lyman break galaxies (the ultraviolet dropouts of 
Steidel et al.~1996) and submillimeter-luminous galaxies
(Smail et al.~1997, Barger et al.~1998, Hughes et al.~1998, Eales et al.~1999).
In addition, a few other high redshift galaxies like F10214+4724 (Rowan-Robinson
et al.~1991) and HR10 (Hu \& Ridgway 1994) have been identified by virtue of
their infrared properties.  The submillimeter-luminous galaxies 
appear to have the same
submillimeter and far-infrared fluxes as would local
ULIGs (Sanders \& Mirabel 1996) if seen at high redshift (Barger et al.~1998).  
 
It therefore is productive to compare the observed optical colors of
the submillimeter-luminous galaxies with our predicted optical colors 
of local ULIGs were they to be seen at high redshift.  This comparison
will help in (i) establishing the connection between local ULIGs and the
submillimeter-luminous galaxies, (ii) identifying candidate ULIGs at high
redshift in color surveys, and (iii) investigating possible links between
the Lyman break galaxies of Steidel et al.~(1996)  and local ULIGs (and by
implication 
between the Lyman-break galaxies and submillimeter-luminous galaxies). 

       The $B - R$ and $B-K$ colors of our sample galaxies placed at different
redshifts are presented in Figure 15.  Elliptical galaxies are shown for
comparison purposes.  Color-color and color-magnitude diagrams
are shown in Figure 16.  Kinks in the curves are artefacts of poor sampling.
The general shapes and positions of the curves are accurate, because the errors
in our SEDs are small.  The forms of the color-color curves are easily 
understood in terms of the shapes of the SEDs.  For example, consider the curve
for IRAS F12112+0305 in the upper panel of Figure 16.  Initially, the gradient
of the curve is shallow: the $B$-band fluxes drop rapidly with increasing $z$
due to the drop of the SED between optical and ultraviolet wavelengths, but the
$R$ and $K$ measurements are still probing a flat part of the SED and so have 
similar $K$-corrections.  When the $R$-band measurements begin to probe 
wavelengths corresponding to the optical-UV drop, the $B - R$ color remains 
approximately constant, but the $B - K$ color continues to redden.  Therefore 
the curves turn upward in the color-color diagram.  Eventually, the $B$-band
measurements begin to probe a flat part of the SED (between the F220W and F140W
measurements), while the $R$-band measurements continue to probe wavelengths in
the region of the optical-UV drop.  The curves then become horizontal: the
$B - R$ color becomes bluer, but the $B - K$ color remains approximately 
constant.  The IRAS F22491$-$1808 curve is somewhat different: the optical-UV 
drop is sufficiently shallow that the bandwidth part of the $K$ correction ---
the $(1 + z)$ term inside the brackets in equations (3) and (4) --- is 
proportionally more important.

      The figures suggests that only one of our galaxies (VII{\ts}Zw{\ts}031) can
appear
redder in $B - R$ than typical elliptical galaxies at $z = 0.8$.  This is the
redshift at which elliptical galaxies are reddest -- see Coleman et al.~(1980). 
The $B - K$ colors of high-$z$ ellipticals are redder than those of all our 
sample galaxies at all redshifts.  
It is therefore
clear that we cannot hope to obtain complete samples of ULIGs at
high-$z$ using color surveys alone.
For example, we could 
not have found the hyperluminous high-$z$ galaxy IRAS F10214+4724 this way.  
Nevertheless the reddest ULIGs should easily be detectable at $z \geq
1$, as they would have colors redder than any normal galaxies.  
For example, very red faint galaxies have been found by
Hu \& Ridgway (1994), one of which (HR10)
has since been detected by SCUBA
(Cimatti et al.~1999).

A comparison with the $V-I$ colors of the SCUBA sources detected by
Smail et al.~(1997, 1998) is given in Figure 17.
We use $V-I$ colors as opposed to the $B-K$ colors presented in Figure 16
because $K$-band colors are not yet available for a significant number
of SCUBA sources.
Note that the Smail et al.~(1998) galaxies are seen through low redshift
clusters and so are magnified by as much as a factor of $\sim 2 - 5$
through gravitational lensing.  This means that the we need to shift the points
in Figure 17 horizontally to the
right by up to one magnitude when comparing the
points with the lines. 
The $V-I$ colors are only
weakly affected by gravitational lensing, if at all.

We see that VII Zw 031 at high redshift would appear redder in $V-I$ than
any of the Smail et al.~galaxies; on the other hand all the Smail et
al.~galaxies 
(except the brightest two, which are cD galaxies) have $V-I$
colors close (within one magnitude)
to those of IRAS F12112+0305 and IRAS F22491$-$1808 seen at high redshift.   
The high-redshift Lyman break galaxies, recently discovered
by Steidel et al.~(1996), are also systematically bluer (in $R-K$)
than our sample galaxies would
appear if seen at the same redshift (see Table 7). 
We use $R-K$ colors for this comparison (as opposed to $B-K$) because
our measurements do not probe far enough into the ultraviolet to permit
us to predict $B$ magnitudes for our sample galaxies placed at the very
high redshifts of the sample of Lyman break galaxies used in Table 7.

One implication of these results is that current studies of large 
optically-selected samples of
field galaxies that attempt to determine cosmological star formation rates
(e.g.~Madau et al.~1996) will be missing very red ULIGs 
like VII Zw 031.
Furthermore, if star formation rates 
are derived solely on the basis of rest-frame
ultraviolet colors, they may then be incorrectly determined (Hughes
et al.~1998) for 
galaxies like IRAS F12112+0305 and IRAS F22491$-$1808, which have optical
colors of normal galaxies seen at high redshift.
If ULIGs are a significant population at high redshift, as they
appear to be (Smail et al.~1997, Barger et al.~1998, Hughes et al.~1998,
Eales et al.~1999), then these optical 
studies will clearly underestimate the star formation rate
of the universe at high redshift. 
ore recently, the need to correct optically-derived star formation rates
for dust extinction has been recognized (Calzetti 1999, Pettini et al.~1999).
Deep submillimeter surveys like those currently being carried out with SCUBA
will in the future determine how necessary such corrections are
and whether the corrections mostly come from a small amount of extinction in
average galaxies or from a few extreme galaxies being misidentified or
issed from the
samples altogether.

\section{SUMMARY}

      We have presented {\it HST\/} FOC images of three low redshift ULIGs and
used the
measured fluxes to compute SEDs for the galaxies.  VII{\ts}Zw{\ts}031, IRAS
F12112+0305,
IRAS F22491$-$1808) were selected to span the range of properties of ULIGs;
they have very different ultraviolet SEDs.  All have SEDs that drop between 
optical and ultraviolet wavelengths, but the drops range from a factor of $\sim
3$ in IRAS F22491$-$1808 to a factor of $\sim 100$ in VII{\ts}Zw{\ts}031.  We
interpret
this as being due to different internal extinctions in the galaxies.  A high
ultraviolet extinction in VII{\ts}Zw{\ts}031 is also suggested by extrapolating
the 
optical internal extinction measured by Djorgovski et al.~(1990).  
Implications of our results are as follows:

\begin{enumerate}

\item
Only one of our three galaxies (VII{\ts}Zw{\ts}031) has an SED that drops
steeply enough
to ultraviolet wavelengths that its $B - R$ color as seen at high $z$ is redder
than the $B - R$ colors of elliptical galaxies seen at any $z$.  
The other two galaxies have $B - R$ colors that are normal for faint field
galaxies.
Therefore only the reddest ULIGs can be found at high $z$ on the basis of
optical and near-infrared
colors alone.  The implication is then that submillimeter surveys, like those
with
SCUBA, are required to obtain complete samples of ULIGs at high $z$.

\item
Galaxies like VII{\ts}Zw{\ts}031 seen
at high redshift may not appear in any current optically selected field
galaxy sample because their rest-frame UV fluxes are too attenuated.
Galaxies like IRAS F12112+0305 and IRAS F22491$-$1808 would appear in
these samples, but their star-formation rates, if derived solely from their
rest-frame ultraviolet fluxes, may be serious underestimates.
Therefore, any conclusions about the star formation rate of the Universe that 
are derived from optically selected samples are subject to the caveat that ULIGs 
may be missing or not correctly identified. 
The submillimeter surveys currently in progress will tell us how
serious this worry is.

\item
VII{\ts}Zw{\ts}031 seen at high redshift would have a redder $V-I$
colour than any of the SCUBA sources studied at optical wavelengths
by Smail et al.~(1998).  However, most of the Smail et al.~galaxies have
similar $V-I$ colors to either IRAS F12112$-$0305 or
IRAS F22491$-$1808 seen at $z>1$.  
The Lyman break galaxies in the sample of Pettini et al.~(1998)
have $R-K$ colors that are systematically
bluer than those of our sample galaxies
if seen at the same redshift. 

\end{enumerate}

\acknowledgments
It is a pleasure to acknowledge useful discussions with Bill Vacca, Jeff 
Goldader, Esther Hu, Joseph Jensen, Robert Joseph, Joseph Mazzarella,
and Andrew Blain.  This
research has made use of the NASA/IPAC extragalactic database (NED), which is 
operated by the Jet Propulsion Laboratory, Caltech, under agreement with the 
National Aeronautics and Space Administration.  JK and NT were supported in 
part by {\it HST\/} grant GO-3913.01-91A.  DBS was supported in part by NASA 
grant NAGW-3938.

\clearpage

\centerline{FIGURE CAPTIONS}

\vskip 0.3in

\figcaption[Fig1.eps]{Ground-based R-band images of our sample galaxies.  Each
field is
$51^{\prime\prime} \times 51^{\prime\prime}$; this corresponds to 49 $\times$ 49
kpc for VII{\ts}Zw{\ts}031, 63 $\times$ 63 kpc for IRAS F12112+0305, and 67
$\times$ 67
kpc for IRAS F22491$-$1808.  All images in this paper have north up and east to
the left.  The VII{\ts}Zw{\ts}031 image is from the data of 
Mazzarella et al.~(1999); the
other
images are from the University of Hawaii 2.24 m telescope (see \S{\ts}4).}

\figcaption[Fig2.eps]{The {\it HST\/} FOC image of VII{\ts}Zw{\ts}031 taken
through the
F220W filter.  The optical center of the galaxy is identified with a $+$.  The
figure shows a section of the image about 7\sd3 = 7 kpc on a side.  The left 
panel shows the raw data.  The other two panels show deconvolved data.  For all
of the deconvolutions, the point-spread functions (PSF) were generated using
the {\it HST\/} TinyTim software for a star with $B-V = 1.1$.  The center panel
shows the data after 20 iterations of Lucy (1974) -- Richardson (1972) 
deconvolution.  The right panel shows the data after maximum
entropy deconvolution using the STSDAS MEMSYS-5 software with the MEM front end
(N. Weir 1991, private communication).  It is the mean of two images, which are
the raw data deconvolved with the PSF described above and using intrinsic 
correlation functions of correlation length 0\sd02 and 0\sd16 (the PSF has FWHM
$\simeq$ 0\sd1).  An explanation of the meaning and use of intrinsic correlation
functions and a discussion of the advantages of the above approach are given in 
detail in the MEM manual.}

\figcaption[Fig3.eps]{Growth curves for the F220W image of VII{\ts}Zw{\ts}031
based on
aperture photometry carried out as described in the text.  The sky values for 
the six growth curves are 2170, 2190, 2210, 2230, and 2250  
counts arcsec$^{-2}$
from the top down. 
The heavy line plateaus over the widest range in radius; we adopt it and use the
others to estimate the errors.  The resulting total brightness of
VII{\ts}Zw{\ts}031 at
2300 \AA~is $9400 \pm 2000$ counts.}

\figcaption[Fig4.eps]{Growth curves for the F140W image of VII{\ts}Zw{\ts}031. 
The 
corresponding sky values are 
5170, 5210, 5250, 5290, and 5330 
counts arcsec$^{-2}$ from the top down.}

\figcaption[Fig5.eps]{The F220W image of IRAS F12112+0305.  The box is a square
of side 9\sd1 = 12.1 kpc.  The two crosses mark the centers of the two galaxies
that make up this interacting system.  The fingerprint pattern introduced by the
geometric distortion correction is visible at a low level in this image.}

\figcaption[Fig6.eps]{Growth curves for the F220W image of IRAS F12112+0305. The
corresponding sky values are 1550, 1570, 1590,
1610, 1630, and 1650 counts arcsec$^{-2}$ from the top down.}

\figcaption[Fig7.eps]{Growth curves for the F140W image of IRAS F12112+0305. The
corresponding sky values are 
4545, 4590, 4630, 4650, 4670, 4710, and 4755 
counts arcsec$^{-2}$ from the top down.}

\figcaption[Fig8.eps]{(a) The F220W image of IRAS F22491$-$1808.  The box
is a square of side 11.3 arcseconds = 16.0 kpc.  The $+$ marks the optical 
center of the galaxy; (b) the F140W image of IRAS F22491$-$1808.  The
scale and orientation are the same as in (a).}

\figcaption[Fig9.eps]{Growth curves for the F220W image of IRAS F22491$-$1808.
Rectangular apertures were used as described in the text; $r$ represents the 
distance from the center to the edge of the aperture in the direction orthogonal
to that for which large-scale fluctuations in the detector are seen.  The width
of the detector in that direction is then $(2 r + 1)$ arcseconds.  The width
of the aperture in the orthogonal direction (i.{\ts}e., that of the 
fluctuations) is fixed at 8\sd9.  The sky values are 166, 176, 186, 196, and 
206 counts arcsec$^{-2}$ from the top down.}

\figcaption[Fig10.eps]{Growth curves for the F140W image of IRAS F22491--1808. 
Radius has the same meaning as in Figure 9.  The sky values are 920, 940, and 
960 counts arcsec$^{-2}$ from the top down.}

\figcaption[Fig11.eps]{The rest-frame spectral energy distribution of
VII{\ts}Zw{\ts}031.
The open symbols represent the {\it IRAS\/} 12, 25, 60, and 100 \micron\ 
measurements (the uncertainties in the {\it IRAS} measurements are smaller than
the 
symbols) and the ground-based $g$, $r$, $i$, and $K$-band measurements.  The
filled symbols and error bars are the {\it HST\/} FOC results.}

\figcaption[Fig12.eps]{The rest-frame spectral energy distribution of IRAS
F12112$+$0305.  Open and closed symbols have the same meaning as in Figure 11.
The optical and near-infrared bandpasses are $B$, $V$, $R$, $I$, and 
$K^{\prime}$.}

\figcaption[Fig13.eps]{The rest-frame spectral energy distribution of IRAS 
F22491--1808.  Open and closed symbols have the same meaning as in Figure 11.
The optical and near-infrared bandpasses are $B$, $V$, $R$, $I$, and $K$.}

\figcaption[Fig14.eps]{The 
SEDs of our sample galaxies, all 
(arbitrarily) normalized to have the same luminosity 
as F10214+4724 at $\times$ 10$^{15}$ Hz.  The symbols have the same meanings as
In Figures 11 to 13.  The error bars for all the points are given in
Figures 11 to 13; note that the bluemost point for VII{\ts}Zw{\ts}031 represents
only
a 0.45$\sigma$ detection.}

\figcaption[Fig15.eps]{The $B - R$ and $B - K$ colors of our sample galaxies as
a function of redshift.  These were computed using the $K$ corrections in Table 
6.  The colors of $M_B = -20$ elliptical galaxies are taken from the 
measurements of Coleman et al.~(1980).  We use the near-infrared $K$ corrections
for elliptical galaxies of Huang et al.~(1997).}

\figcaption[Fig16.eps]{The upper panel is a color-color diagram showing how our
sample galaxies would appear at different redshifts.  The elliptical galaxy is
assumed to have $M_B = -20$ at $z = 0$.  Each of the curves is labelled with two
numbers.  The end labelled ``0'' represents the galaxy at $z = 0$.  We present
each curve to the highest redshift that is constrained by our SEDs; this 
redshift is the label at the other end of the curve.  We use F140W data for IRAS
F12112+0305 and IRAS F22491$-$1808 only.  The lower panel is a corresponding 
color-magnitude diagram.  The curves and symbols have the same meaning as in the
upper panel.  Elliptical galaxies are not included on this panel; they can
occupy almost any region of this plot, depending on their absolute magnitude
and redshift. 
In both panels,
IRAS F10214+4724 (Rowan-Robinson et al.~1993) is indicated as a cross.} 

\figcaption[Fig17.eps]{The $V-I$ versus $I$ color-magnitude diagram
for our sample galaxies seen at different redshifts.
The curves have the same meanings as in the lower panel of Figure 16.
The filled circles represent the colors and magnitudes for the sample
of SCUBA sources studied by Smail et al.~(1998).}  


\clearpage

\begin{references}

\reference{} Barger, A.~J., Cowie, L.~L., Sanders, D.~B., Fulton, E., Taniguchi,
Y., 
Sato, Y., Kawara, K., \& Okuda, H.~1998,
Nature, 394, 298 

\reference{} Baum, S.~(ed.)~1994, HST Data Handbook (Baltimore: Space Telescope
      Science Institute)

\reference{b1} Baxter, D.~A.~1993, in Calibrating Hubble Space Telescope, 
      ed.~J.~C.~Blades \& S.~J.~Osmer (Baltimore: Space Telescope Science
      Institute), 116 

\reference{b2} Blumenthal, G.~R., Faber, S.~M., Primack, J.~R., \& Rees,
      M.~J.~1984, Nature, 311, 517

\reference{b3} Burstein, D., \& Heiles, C.~1982, AJ, 87, 1165

\reference{} Calzetti, D.~1999, in 
Dwarf galaxies 
and cosmology: 
Proc.~XVIII Moriond meeting, ed.~T.~X.~Thuan et al.~(Gif-sur-Yvette:
Editions Fronti\`eres), in press 

\reference{c1} Cardelli, J.~A., Clayton, G.~C., \& Mathis, J.~S.~1989, ApJ, 
      345, 245

\reference{} Carlberg, R.~G.~1986, ApJ, 310, 593

\reference{} Cimatti, A., Andreani, P., Rottgering, H., Tilanus,
R.~1999, in Proc.~X Rencontres de Blois, ed.~B.~Guiderdoni
et al.~(Gif-sur-Yvette: Editions Fronti\`eres), in press

\reference{c2} Clements, D.~L., Sutherland, W.~J., McMahon, R.~G., \& Saunders,
      W.~1996, MNRAS, 279, 477

\reference{c3} Coleman, G.~D., Wu, C.~C., \& Weedman, D.~W.~1980, ApJS, 43, 393

\reference{} Conti, P.~S., \& Vacca, W.~D.~1994, ApJ, 423, L97

\reference{d1} de Grijp, M.~H.~K., Miley, G.~K., Lub, J., \& de Jong, T.~1985, 
      Nature, 314, 240

\reference{d2} Djorgovski, S., De Carvalho, R.~R., \& Thompson, D.~J.~1990, AJ, 
      99, 1414

\reference{} Doyon, R., Wells, M., Wright, G.~S., Joseph, R.~D., Nadeau, D., \&
      James, P.~A.~1994, ApJ, 437, L23

\reference{} Eales, S.~A., Lilly, S.~J., Gear, W.~K., Bond, J.~R., Dunne, L.,
Hammer, F., 
Le F\`evre, O., \& Crampton, D.~1999, ApJ, in press  
(astro-ph/9808040)

\reference{e1} Eisenhardt, P.~R., Armus, L., Hogg, D.~W., Soifer, B.~T., 
      Neugebauer G., \& Werner, M.~W.~1996, ApJ, 461, 72 

\reference{e2} Elias, J.~H., Frogel, J.~A., Matthews, K., \& Neugebauer, 
      G.~1982, AJ, 87, 1029

\reference{f1} Freeman, K.~C.~1970, ApJ, 160, 811

\reference{} Gunn, J.~E.~1987, in Nearly Normal Galaxies: From the Planck Time
      to the Present, ed.~S.~M.~Faber (New York: Springer-Verlag), 455

\reference{} Hu, E.~M., \& Ridgway, S.~E.~1994, AJ, 107, 1303

\reference{} Huang, J.~S., Cowie, L.~L., Gardner J.~P., Hu, E.~M.,
Songaila, A., \& Wainscoat, R.~J.~1997, ApJ, 476, 12 

\reference{} Hughes, D.~1998, Nature, 394, 241 

\reference{j1} Jarvis, J.~F., \& Tyson, J.~A.~1981, AJ, 86, 476 

\reference{j2} Joseph, R.~D., \& Wright, G.~S.~1985, MNRAS, 214, 87

\reference{k1} Kinney, A.~L., Calzetti, D., Bica, E., \& Storchi-Bergman,
      T.~1994, ApJ, 429, 172

\reference{} Kormendy, J.~1989, ApJ, 342, L63

\reference{} Kormendy, J.~1990, in Dynamics and Interactions of Galaxies, 
      ed.~R.~Wielen (New York: Springer-Verlag), 499

\reference{} Kormendy, J., \& Djorgovski, S.~1989, ARA\&A, 27, 235

\reference{k2} Kormendy, J., \& Sanders, D.~B.~1992, ApJ, 390, L53

\reference{l1} Landolt, A.~U.~1992, AJ, 104, 340

\reference{} Lonsdale, C.~J., Diamond, P.~J., Smith, H.~E., \& Lonsdale,
C.~J.~1994, Nature, 370, 117  

\reference{} Lucy, L.~B.~1974, AJ, 79, 745

\reference{} Madau, P., Ferguson, H.~C., Dickinson,
.~E., Giavalisco, M., Steidel, C.~C., \& Fruchter, A.~1996, MNRAS,
283, 1388

\reference{m2} Mathis, J.~S.~1990, ARA\&A, 28, 37

\reference{} Mazzarella, J.~M.~et al.~1999, in preparation

\reference{m1} Meurer, G.~R., Heckman, T.~M., Leitherer, C., Kinney, A., Robert,
      C., \& Garnett, D.~R.~1995, AJ, 110, 2665

\reference{m3} Moshir, M., Kopan, G., Conrow, J., McCallon, H., \& Hacking,
      P.~1992, Explanatory Supplement to the $IRAS\/$ Faint Source Survey 
      Version 2 (Pasadena: JPL)

\reference{m4} Murphy, T.~W., Armus, L., Matthews, K., Soifer, B.~T., \&
      Mazzarella, J.~M.~1996, AJ, 111, 1025

\reference{} O'Connell, R.~W., Gallagher, J.~S., Hunter, D.~A., \& Colley, 
      W.~N.~1995, ApJ, 446, L1

\reference{o1} Oliver, S., Rowan-Robinson, M., Broadhurst, T.~J., McMahon,
      R.~G., \& Saunders, W.~1996, MNRAS, 280, 673

\reference{} Ostriker, J.~P.~1980, Comments Ap, 8, 177

\reference{p1} Perault, M.~1987, PhD thesis, University of Paris

\reference{} Pettini, M., Kellogg M., Steidel, C.~C.,
Dickinson M., Adelberger K.~L., \& Giavalisco, M.~1998, ApJ, 508, 539  

\reference{} Pettini, M., Steidel, C.~C., 
Adelberger, K.~L., Kellogg, M., Dickinson, M.,
\& Giavalisco, M.~1999, in 
Cosmic Origins: evolution of galaxies, stars, planets, and life,
ed.~J.~M.~Shull, C.~E.~Woodward, \& H.~A.~Thronson 
(San Francisco: ASP), in press (astro-ph/9808117)  

\reference{} Richardson, W.~H.~1972, J.~Opt.~Soc.~America, 62, 52

\reference{r1} Rowan-Robinson, M.,~et al.~1991, Nature, 351, 719 

\reference{r2} Rowan-Robinson, M.,~et al.~1993, MNRAS, 261, 513 

\reference{s1} Sage, L., \& Solomon, P.~1987, ApJ, 321, L103

\reference{} Sandage, A.~1995, in The Deep Universe, Saas-Fee Advanced Course
      23, ed.~B.~Binggeli \& R.~Buser (New York: Springer-Verlag), 1

\reference{s2} Sanders, D.~B., \& Mirabel, I.~F.~1996, ARA\&A, 34, 749

\reference{s3} Sanders, D.~B., Scoville, N.~Z., \& Soifer, B.~T.~1991, ApJ,
      370, 158

\reference{s4} Sanders, D.~B., Soifer, B.~T., Elias, J.~H., Madore, B.~F., 
      Matthews, K., Neugebauer, G., \& Scoville, N.~Z.~1988a, ApJ, 325, 74

\reference{s5} Sanders, D.~B., Soifer, B.~T., Elias, J.~H., Neugebauer, G.,
     Matthews, K.~1988b, ApJ, 328, L35  

\reference{s6} Sanders, D.~B.,~et al.~1999, in preparation
 
\reference{} Schechter, P.~1976, ApJ, 203, 297

\reference{} Schweizer, F.~1996, AJ, 111, 109

\reference{} Schweizer, F., Miller, B.~W., Whitmore, B.~C., \& Fall, S.~M.~1996,
      AJ, 112, 1839

\reference{} Scoville, N.~Z., Yun, M.~S., \& Bryant, P.~M.~1997, ApJ, 484,
702 
 
\reference{} Shaya, E.~J., Dowling, D.~M., Currie, D.~G., Faber, S.~M., \&
      Groth, E.~J.~1994, AJ, 107, 1675

\reference{} Smail, I., Ivison, R.~J., \& Blain, A.~W.~1997, 
ApJ, 490, L5

\reference{} Smail, I., Ivison, R.~J., Blain, A.~W., \& Kneib, J.~-P.~1998, 
ApJ, 507, L21

\reference{s10} Solomon, P.~M., Downes, D., Radford, S.~J.~E., \& Barrett,
      J.~W.~1997, ApJ, 478, 144  

\reference{} Steidel, C.~C., \& Hamilton, D.~1992, AJ, 104, 941

\reference{} Steidel, C.~C., Giavalisco, M., Dickinson, M.,
\& Adelberger, K.~1996, AJ, 112, 352 

\reference{s11} Stetson, P.~B.~1987, PASP, 99, 191

\reference{} Tacconi-Garman, L.~E., Sternberg, A., \& Eckart, A.~1997, preprint

\reference{t1} Thuan, T.~X., \& Gunn, J.~E.~1976, PASP, 88, 543

\reference{t2} Trentham, N.~1997, A\&A, 321, 81 

\reference{w1} Wade, R.~A., Hoessel, J.~G., Elias, J.~H., \& Huchra, J.~P.~1979,
      PASP, 91, 35

\reference{w2} Wainscoat, R.~J., \& Cowie, L.~L.~1992, AJ, 103, 332

\reference{} Whitmore, B.~C., \& Schweizer, F.~1995, AJ, 109, 960

\reference{} Whitmore, B.~C., Schweizer, F., Leitherer, C., Borne, K., 
      \& Robert, C.~1993, AJ, 106, 1354

\end{references}
\end{document}